\documentclass[useAMS,usenatbib]{mnras}
\usepackage{rotating}
\usepackage{lscape}
\usepackage{longtable}
\usepackage{lastpage}

\hypersetup{draft}

\defcitealias{jenkins11}{Jenkins et al. 2011a} 
\defcitealias{jenkins11a}{Jenkins et al. (2011a)} 	
\defcitealias{jenkins11b}{Jenkins et al. 2011b} 
\defcitealias{jenkins13a}{Jenkins et al. 2013a} 
\defcitealias{jenkins2013a}{Jenkins et al. (2013a)} 	
\defcitealias{jenkins13b}{Jenkins et al. 2013b} 	
\defcitealias{jenkins13c}{Jenkins et al. 2013c}		
\defcitealias{jenkins14}{Jenkins et al. 2014a}

\def\ms{ms$^{-1}$}

\def\me{$M_{\rm{\oplus}}$}
\def\re{$R_{\rm{\oplus}}$}

\def\rsun{$R_{\odot}$}

\DeclareSymbolFont{UPM}{U}{eur}{m}{n}
\DeclareMathSymbol{\umu}{0}{UPM}{"16}
\let\oldumu=\umu
\renewcommand\umu{\ifmmode\oldumu\else$\oldumu$\fi}
\newcommand\micro{\umu}
\def\micron{\micro m}
\let\microns \micron

\title[No transit for Proxima b]{Proxima Centauri b is not a transiting exoplanet}

\author[J.S. Jenkins et al.]{James S.\ Jenkins$^{1,2,\ast}$,
Joseph Harrington$^{3}$,
Ryan C.\ Challener$^{3}$,
Nicol\'as T. Kurtovic$^{1}$, \newauthor
Ricardo Ramirez$^{1}$,
Jose Pe{\~n}a$^{1}$,
Kathleen J. McIntyre$^{3}$,
Michael D. Himes$^{3}$, \newauthor
Eloy Rodr\'iguez$^{4}$,
Guillem Anglada-Escud\'e$^{5}$,
Stefan Dreizler$^{6}$,
Aviv Ofir$^{7}$, \newauthor
Pablo A. Pe{\~n}a Rojas$^{1}$,
Ignasi Ribas$^{8,9}$, 
Patricio Rojo$^{1}$,
David Kipping$^{10}$, \newauthor
R. Paul Butler$^{11}$, 
Pedro J. Amado$^{4}$, 
Cristina Rodr\'iguez-L\'opez$^{4}$, \newauthor
Eliza M.-R.\ Kempton$^{12,13}$, 
Enric Palle$^{14,15}$,
Felipe Murgas$^{14,15}$ \\
$^1$Departamento de Astronomía, Universidad de Chile, Camino El Observatorio 1515, Las Condes, Santiago, Chile \\
$^2$Centro de Astrof\'isica y Tecnolog\'ias Afines (CATA), Casilla 36-D, Santiago, Chile \\
$^3$Planetary Sciences Group, Department of Physics, University of Central Florida, Orlando, Florida, USA \\
$^4$Instituto de Astrof\'isica de Andaluc\'ia (IAA, CSIC)
Glorieta de la Astronom\'ia, s/n E-18008 Granada, Spain \\
$^5$School of Physics and Astronomy, Queen Mary University of London, 327 Mile End Road, London E1 4NS, UK \\
$^6$Institut f\"ur Astrophysik, Georg-August-Universit\"at G\"ottingen Friedrich-Hund-Platz 1, 37077 G\"ottingen, Germany \\
$^7$Department of Earth and Planetary Sciences, Weizmann Institute of Science, 234 Herzl Street, Rehovot 76100, Israel \\
$^8$Institut de Ci\'encies de l'Espai (ICE, CSIC), C/Can Magrans s/n, Campus UAB, 08193 Bellaterra, Spain \\
$^9$Institut d'Estudis Espacials de Catalunya (IEEC), 08034 Barcelona, Spain \\
$^{10}$Department of Astronomy, Columbia University, 550 W 120th Street, New York NY 10027 \\
$^{11}$Department of Terrestrial Magnetism, Carnegie Institution of Washington, 5241 Broad Branch Road NW, Washington D.C. USA 20015-1305 \\
$^{12}$Department of Physics, Grinnell College, 1116 8th Ave., Grinnell, IA 50112, USA \\
$^{13}$Department of Astronomy, University of Maryland, College Park, MD 20742, USA \\
$^{14}$Instituto de Astrof\'{i}sica de Canarias (IAC), E-38205 La Laguna, Tenerife, Spain \\
$^{15}$Departamento de Astrof\'{i}sica, Universidad de La Laguna (ULL), E-38206 La Laguna, Tenerife, Spain}
\begin{document}

\date{Submitted 2019 March 28; Accepted 2019 April 30}

\pagerange{\pageref{firstpage}--\pageref{lastpage}} \pubyear{2019}

\maketitle

\label{firstpage}

\begin{abstract}

We report Spitzer Space Telescope observations during predicted transits of the exoplanet Proxima Centauri b.  As the nearest terrestrial habitable-zone planet we will ever discover, any potential transit of Proxima b would place strong constraints on its radius, bulk density, and atmosphere. Subsequent transmission spectroscopy and secondary-eclipse measurements could then probe the atmospheric chemistry, physical processes, and orbit, including a search for biosignatures.  However, our photometric results rule out planetary transits at the 200~ppm level at 4.5~{\microns}, yielding a 3$\sigma$ upper radius limit of 0.4~{\re} (Earth radii).  Previous claims of possible transits from optical ground- and space-based photometry were likely correlated noise in the data from Proxima Centauri's frequent flaring.  Follow-up observations should focus on planetary radio emission, phase curves, and direct imaging.  Our study indicates dramatically reduced stellar activity at near-to-mid infrared wavelengths, compared to the optical.  Proxima b is an ideal target for space-based infrared telescopes, if their instruments can be configured to handle Proxima's brightness.

\end{abstract}


\section{Introduction}

The search for the nearest small planets has accelerated in recent years with the development of purpose-built instrumentation \citep[e.g.,][amongst others]{mayor03,crane06,cosentino12,pepe13,QuirrenbachEtal2018spieCARMENES}.  Some highlights include the multi-planet systems orbiting the nearby stars HD~69830 \citep{lovis06}, HD~10180 \citep{lovis11,tuomi12}, HD~40307 \citep{tuomi13}, or 61 Virginis \citep{vogt10}.  The value of these nearby planetary systems significantly increases when the planets are found to transit their host stars, like those orbiting HD~219134 \citep[see for example,][]{vogt15,motalebi15,gillon17b}, 55 Cancri e \citep{winn11}, or the recent Transiting Exoplanet Survey Satellite discovery of $\Pi$~Mensae~c \citep{huang18}.

M-dwarf stars are particularly fruitful targets.  Their occurrence rate for planets with masses below 10~{\me} (Earth masses) is at least one per star, with a habitable zone (HZ) occurrence rate in the mass range 3--10~{\me} of 0.21$^{+0.03}_{-0.05}$ planets per star \citep{tuomi14}.  Within this population are dense multi-planet systems like the seven planet candidates orbiting GJ~667C \citep{AngladaEscudeEtal2013aaGJ667C} and the four planets around GJ~876 \citep{rivera10,jenkins14}. Furthermore,  \cite{RibasEtal2018natBarnardb} detected a small planetary candidate orbiting Barnard's star.

M dwarfs host some spectacular transiting systems, particularly for the low-mass population, both inside and outside of the HZ.  The super-Earth GJ~1214~b transits its host star \citep{charbonneau09}, allowing studies of its atmospheric composition \citep[e.g.,][]{bean10,rackham17}. The star LHS~1140 hosts two transiting planets \citep{dittmann17,ment18}.  But, the standard-bearer in this class is the TRAPPIST-1 system, with seven small transiting exoplanets, at least three of which are in the HZ \citep{gillon17a}.

Proxima Centauri~b \citep[hereafter Proxima~b]{AngladaEscudeEtal2016natProxCenb} provides potentially the best possible opportunity for exoplanet characterization.  Not only does the planet orbit our nearest stellar neighbour, but it has an equilibrium temperature and minimum mass similar to the Earth, meaning it could be rocky and have liquid surface water.  If it transits, characterisation of its atmosphere and surface would be possible. \cite{KippingEtal2017ajNoProxMOST} used optical photometry from the Microvariability and Oscillations of STars (MOST) Space Telescope.  Although they reported a candidate signature matching the planet's expected properties, they could not confirm the feature.  \cite{LiEtal2017rnaasProxCandidate} also reported a possible optical transit of Proxima~b using data taken with the 30~cm telescope at Las Campanas, but again without confirmation.  Using the Bright Star Survey Telescope in Antartica, \cite{LiuEtal2018ajProxPrelim} report a number of transit-like events that could be ascribed to Proxima~b, and assuming large transit timing variations (TTV), they could phase with the event reported in \cite{KippingEtal2017ajNoProxMOST}.  On the contrary, \cite{blank18} were unable to confirm any of the previously reported transit events using optical data spanning 11~years, albeit with heterogeneous and non-continuous data sets.  They do, however, confirm the impact of high stellar activity on optical light curves.  

Observations at longer wavelengths mitigate the effects of stellar activity, increasing precision.  In $\S$~\ref{observations} we discuss an observing campaign with the Spitzer Space Telescope at 4.5~{\microns} to search for the Proxima~b's transits.  We then present new constraints on the mass provided by the latest radial-velocity (RV) data in $\S$~\ref{rvanalysis}, both for Proxima~b and any planets interior to its orbit.  Finally, we summarise our findings in $\S$~\ref{conclusions}.

\section{Photometric Observations and Analysis}\label{observations}

\begin{figure*}
	\hspace{-0.7cm}\vspace{-0.3cm}\includegraphics[width=8.8cm,angle=270]{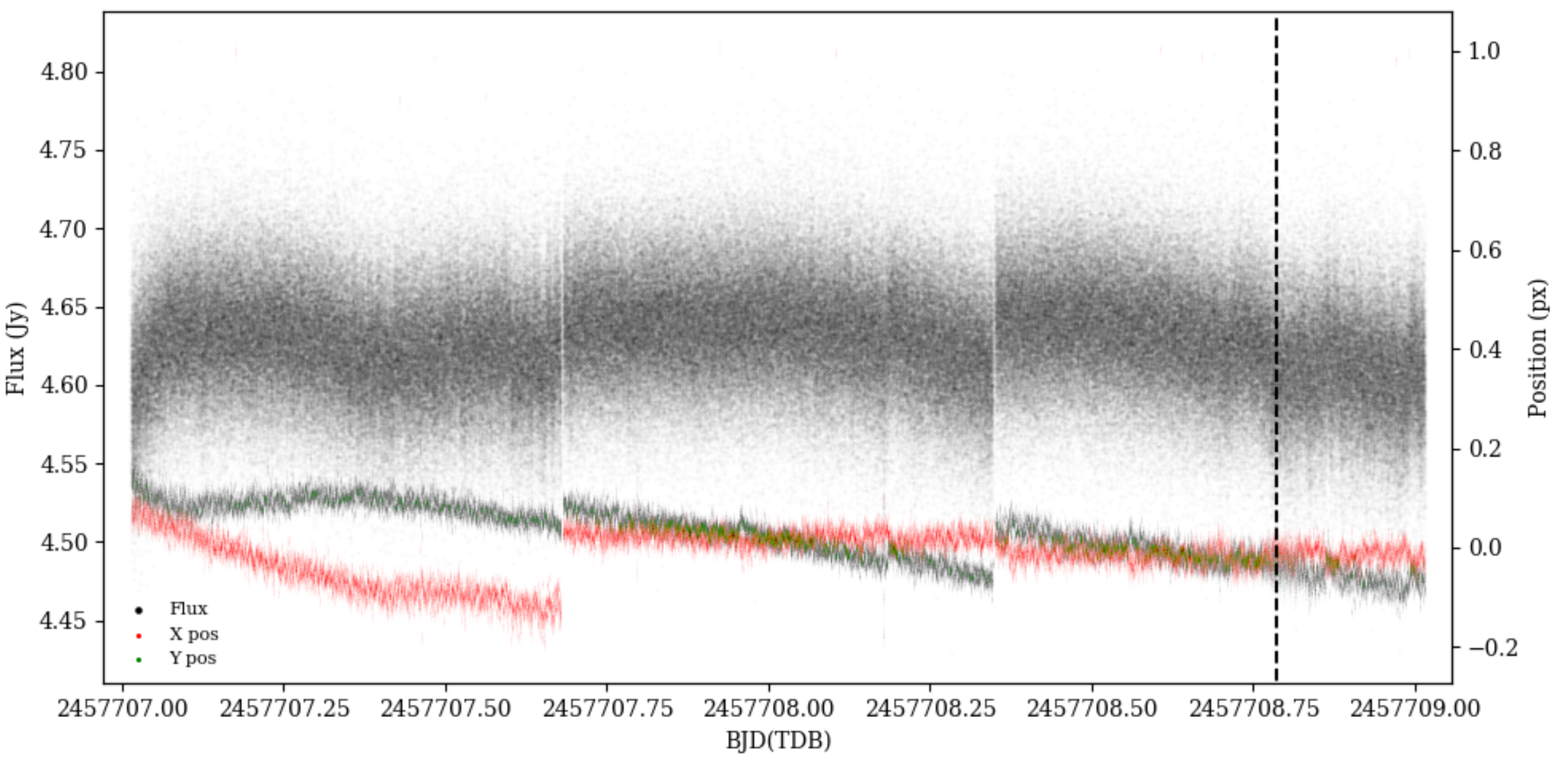}
    \caption{Raw Proxima Centauri photometry (black points) and motion of the stellar point-spread function in $x$ and $y$ (red and green points, respectively), determined using the asymmetric Gaussian fitting method. Breaks in the light curve occur at pointing resets.  The vertical dashed line marks the center of the asymmetric feature.}
    \label{fig:rawlc}
\end{figure*}

In November 2016, we observed Proxima Centauri for over 48~hours in the 4.5~{\micron} band of the InfraRed Array Camera (IRAC, \citealp{FazioEtal2004apjIRAC}) on the Spitzer Space Telescope \citep{WernerEtal2004apjsSpitzer}, with target reacquisition roughly every 16 hours.  We centered the observation on the predicted transit time of 2,457,708.02 $\pm$ 0.33 BJD, calculated from the original orbital solution for Proxima~b published in \citet{AngladaEscudeEtal2016natProxCenb}.  The subarray mode frame time of 0.02~s resulted in $\sim600,000$ individual frames (Spitzer Proposal ID 13155, PI: James Jenkins). Due to on-board data storage limits, there are short gaps between successive sets of 64 subarray images.  The IRAC heater was off for the duration of the stare.

We used Basic Calibrated Data frames from Spitzer pipeline version S19.2.0. We performed twice-iterated 4$\sigma$ bad pixel rejection at every pixel position within each 64-frame set of subarray images to mask cosmic ray hits, and combined these with masks supplied by Spitzer. 

Two groups within our team, RC and JH at UCF and NT, RR, and JJ at U.\ de Chile, analyzed all the data with completely independent codes, obtaining closely similar results.  The UCF group used its Photometry for Orbits, Eclipses, and Transits pipeline (POET; \citealp{StevensonEtal2012apjBLISS, CubillosEtal2014apjTrES1b}), while the Chilean group wrote a new code, in consultation with the UCF group, but not sharing code in either direction.  The codes performed centering, aperture photometry, and light-curve modeling.

We considered Gaussian, center-of-light, and least-asymmetry \citep{LustEtal2014paspLeastAsym} centering, as well as fixed and (at UCF only) variable-aperture \citep{LewisEtal2013apjHATP2b} photometry.  We selected the optimal centering and photometry method by minimizing the standard deviation of normalized residuals (SDNR) and the binned-$\sigma$ $\chi^2$ ($\chi^2_{\rm bin}$, \citealp{DemingEtal2015apjPLD}) of the model. This second metric looks for a broad-bandwidth solution by comparing a curve of SDNR vs.\ bin size to the expected inverse square root.  We find that \textbf{$\chi^2_{\rm bin}$} more sucessfully selects against correlated noise, so we present the results using that selection criterion. The raw photometry and the position of the target on the detector relative to pixel center are shown in Figure \ref{fig:rawlc}.

To remove IRAC's intrapixel sensitivity variations, we applied both BiLinearly Interpolated Subpixel Sensitivity mapping (BLISS; \citealp{StevensonEtal2012apjBLISS}) and Pixel-Level Decorrelation (PLD; \citealp{DemingEtal2015apjPLD}), using independent codes we each developed.  In brief, BLISS iteratively computes a subpixel-resolution sensitivity grid from the light curve.  We account for other effects (transit features, non-flat baselines, etc.) in other model components, all of which fit simultaneously.  The flux is then:

\begin{equation}
\label{eqn:bliss}
F = F_s Tr(t) M(x,y) R(t),
\end{equation}

\noindent
where $F_s$ is the stellar flux, $Tr$ is a transit model \citep[e.g.,][]{MandelAgolEtal2002apjlTransitShape, RappaportEtal2014apjKOI-2700b}, $M$ is the subpixel sensitivity grid, and $R$ is the non-flat baseline (typically linear or quadratic). PLD corrects the same effect by noting that motion of the target on the detector will be correlated with individual pixel flux values, and models the light curve as a weighted sum of the brightest pixels, after normalisation:

\begin{equation}
F = F_s \left(\sum_{i=1}^{n} c_i \hat{P_i} + Tr(t) + R(t)\right),
\label{eqn:pld}
\end{equation}

\noindent
where $i$ denotes each of $n$ pixels, $c_i$ are the weights, and $\hat{P_i}$ are normalized pixel values. Both methods have been used extensively to correct Spitzer photometry \citep[e.g.,][]{StevensonEtal2012apjBLISS, BlecicEtal2014apjWASP43, CubillosEtal2014apjTrES1b, DemingEtal2015apjPLD, BuhlerEtal2016apjHATP13b, WongEtal2016apjWASP19HATP7}.

We fit each model to every combination of centering method, photometry method, and aperture size.  We used fixed apertures with 1.5--4.0 pixel radii in 0.25 pixel increments, and variable apertures with radii from $\sqrt{N}$ to $\sqrt{N}$+2.0 pixels in 0.25 pixel increments. $N$ is the "noise pixel" parameter (\citealp{LewisEtal2013apjHATP2b}; Spitzer IRAC handbook), defined as

\begin{equation}
    \label{eqn:noisepix}
    N = \frac{\left(\sum I(i)\right)^2}{\sum I(i)^2}
\end{equation}

\noindent
where $I(i)$ is the intensity of pixel $i$, and all pixels within the centering aperture are considered. We used a 17x17 pixel box, centered on the pixel containing Proxima Centauri, for centering. We take the combination of centering and photometry that result in the lowest $\chi^2_{\rm bin}$, Gaussian centering with a fixed 2.0 pixel radius aperture, as the best.

POET finds the best-fitting model using least squares.  Since we find that the Spitzer pipeline tends to overestimate uncertainties, we rescale our photometric uncertainties such that the reduced $\chi^2$ of the best fit is 1. For fits with a BLISS map, we set the $x$ and $y$ widths of the subpixel grid equal to the root-mean-square of the point-to-point variation in the $x$ and $y$ positions found from centering. We also require that each subpixel bin contain at least 4 frames. 

We then explore the parameter space using Multi-Core Markov-Chain Monte Carlo (MC$^3$, \citealp{CubillosEtal2017ajMC3}), a Markov-Chain Monte Carlo (MCMC) wrapper, to determine accurate parameter uncertainties. Our Markov chains use DEMCzs, or ``snooker'', a form of differential evolution Markov Chain, to explore the parameter space efficiently \citep{terBraak2006scDEMC, terBraakVrugt2008scDEMCzs}. We run sufficent iterations for all parameters to pass the Gelman \& Rubin convergence test within 1\% of unity \citep{GelmanRubin1992stscConvergence}.

\begin{figure*}
	\hspace{-0.5cm}\vspace{-0.2cm}\includegraphics[width=18cm]{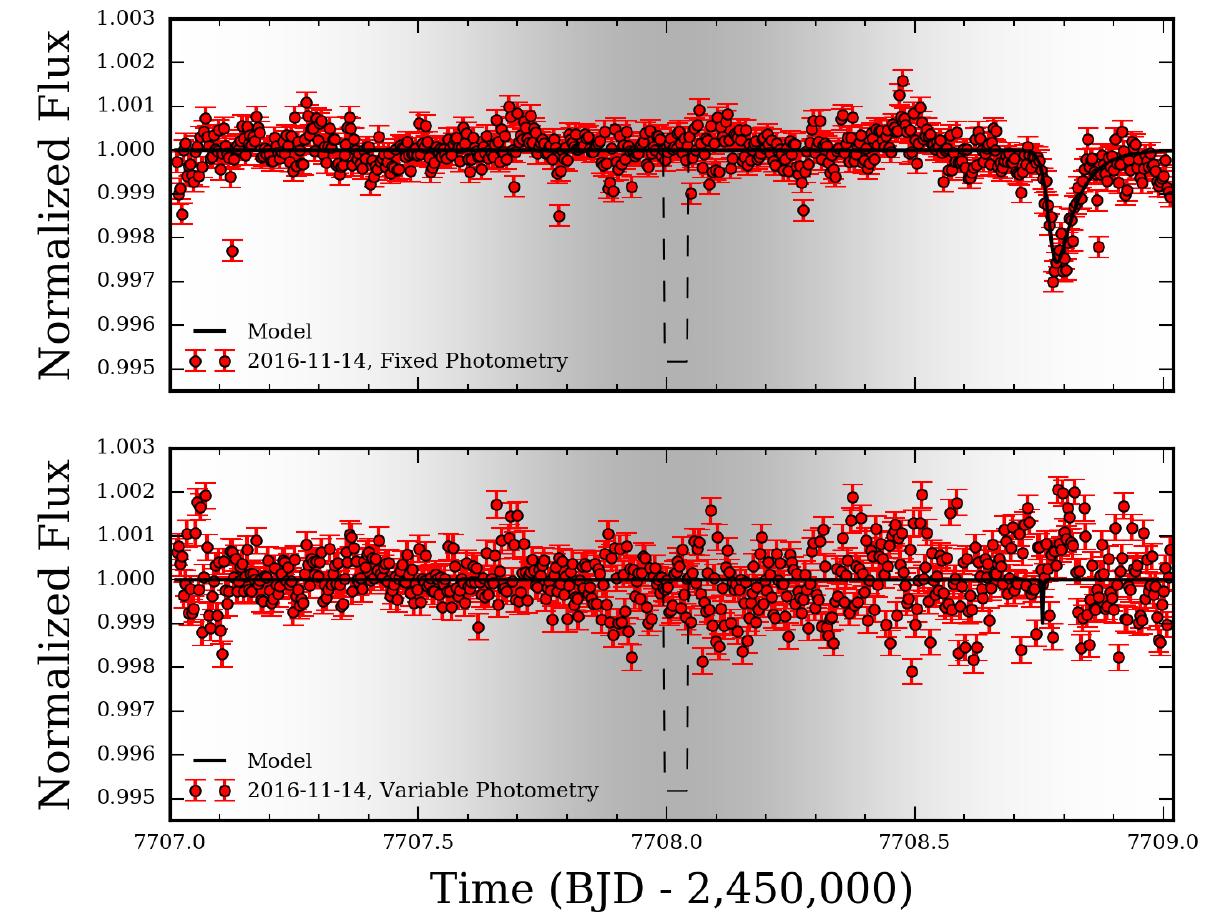}
    \caption{The Spitzer IRAC 4.5~{\microns} light curve for the full 48~hour Proxima Centauri time series.  The dashed line is the predicted Proxima b transit, centered at 2457708.02 $\pm$ 0.33 BJD.  The intensity of the gray-scale background represents the probability of the transit's center from MCMC analysis of the RV data.  The solid black line is an asymmetric hyperbolic secant model, which approximates the feature shape \citep{RappaportEtal2014apjKOI-2700b}. \textbf{Top:} The light curve using Gaussian centering and 2.0 pixel fixed-radius aperture photometry.
    \textbf{Bottom:} The light curve using Gaussian centering and variable-radius aperture photometry (radius $\sqrt{N}$, see Equation \ref{eqn:noisepix}), with no evident asymmetric feature.}
    \label{fig:lc}
\end{figure*}

With this analysis, only a single asymmetric transit-like feature appears, towards the end of the time series (see Fig.~\ref{fig:lc} upper panel).  At $\sim$0.3\% max depth below the continuum, it is smaller than the 0.5\% transit depth that we predict for Proxima b using the parameters determined from the RV modeling effort.  However, if we consider variable-aperture photometry (which is not preferred by our noise-minimization metrics, as this method significantly increases the white noise in the light curve), the feature disappears completely (see Fig.~\ref{fig:lc} lower panel).  The asymmetric transit-like feature corresponds to a telescope vibration that smears the point-spread function, frequently associated with the "noise pixel" parameter. The strength of the vibrational effect depends upon the noise-minimization metric (i.e., the choice of centering and photometry techniques) as well as the decorrelation model (BLISS, PLD, etc.). Further study of this effect, including detection and mitigation techniques, will appear in a forthcoming paper (Challener et al. 2019, in prep).

Proxima flares over 60 times per day \citep{DavenportEtal2016apjlProxFlares}, giving an optical light curve stability at the 0.5--1\% level.  We find an SDNR of 7527 and 9300 ppm for the fixed and variable aperture cases, respectively, in this infrared filter. When binned over a typical $\sim 2$ hour transit, these SDNR drop to 170 and 222 ppm. Taking these as uncertainties and the stellar radius to be 0.154 {\rsun}, we rule out transiting objects with radii $>$0.43 {\re} at the 3$\sigma$ level of confidence, using the more-conservative variable-aperture photometry.  Previously detected features in optical light curves for this star (e.g., \citealp{KippingEtal2017ajNoProxMOST, LiEtal2017rnaasProxCandidate, LiuEtal2018ajProxPrelim}) are not due to Proxima b.  They may be residual correlated noise from the stellar activity.

\section{Spectroscopic Observations and Analysis}\label{rvanalysis}

\subsubsection{The Ultraviolet and Visual Echelle Spectrograph (UVES)}

The RV data from the Ultraviolet and Visual Echelle Spectrograph (UVES) and High Accuracy Radial velocity Planet Searcher (HARPS; see \citealp{AngladaEscudeEtal2016natProxCenb} for details) that were used to discover Proxima b, along with new HARPS data observed as part of the Red Dots\footnote{https://reddots.space/} campaign, allowed further confirmation of the existence of Proxima b, along with improved upper limits on the mass of any additional body orbiting Proxima with an orbit interior to that of Proxima b. The 77 observations from UVES span a baseline of over seven years, with RVs acquired from Julian Date 2,451,634.731 to 2,454,189.714 at signal-to-noise ratios over 100 at 5500\AA, necessary to measure optical RVs at the 1~{\ms} level, and generally taken at low observing cadence (for instance, no measurements were acquired on successive nights).  The resolving power of the UVES spectra was $R=\lambda/\Delta\lambda$ = 100,000 -- 120,000, where $\lambda$ is wavelength, through application of image slicer \#3, which redistributes the light from the $1''$ opening along the 0.3$''$ slit.  An iodine cell placed in the optical light path before entrance into the echelle spectrograph had an operational temperature of 70$^\circ$~C.  It imprinted a dense forest of molecular iodine lines on the stellar spectra between $\sim$5000$\--$6200\AA.  More details on the observational strategy for the UVES spectra can be found in \citet{kurster03}, \citet{endl08}, \citet{zechmeister09}, and \citet{AngladaEscudeEtal2016natProxCenb}.

The treatment of the spectra follow the classical reduction steps for such data \citep[see][]{baranne96,jenkins17}, including debiasing, cosmic ray removal, echelle order location, flatfielding, scattered-light removal, spectral extraction, spectral deblazing, and wavelength calibration.  For UVES, the resulting spectrum, when compared against a previously measured Fourier transform spectrometer (FTS) iodine spectrum at much higher resolution, enables modeling of the iodine spectrum. However, this requires a prior template observation of the star without the iodine cell and without the spectrograph's point-spread function included. The mathematical form of the process is

\begin{equation}
F_o(\lambda) = c [F_t(\lambda) F_i(\lambda + \Delta\lambda)] \ast PSF,
\label{eqn:iod}
\end{equation}

\noindent
where $F_o$ is the observed spectrum, $F_t$ is the observed stellar template without iodine, $F_i$ is the FTS iodine spectrum, $\Delta\lambda$ is the subsequent shift in wavelength between the iodine model and the FTS observation, and $c$ is a normalisation constant. Forward modeling of the iodine spectrum, the stellar spectrum (Proxima), and the instrumental response can yield precise measurements of the star's RV signature. The final UVES internal precision using this technique is $\sim$0.6 m/s.

\subsubsection{High Accuracy Radial velocity Planet Searcher (HARPS)}

The HARPS data cover Julian Dates from 2,457,406.870 to 2,458,027.479, giving rise to 115 measurements after cleaning outliers.  Contrary to the UVES observing cadence, the HARPS data were acquired at very high observing frequency, with multiple observations being taken on individual nights, and many covering successive nights.

HARPS uses a different observing methodology, but there are many similarities in the processing. For initial reduction we used the HARPS Data Reduction Software \citep{MayorEtal2003} to perform the steps outlined above for UVES, extracting calibrated spectra from the raw images. To calculate the RVs, we used the HARPS-TERRA package \citep{AngladaEscudeEtal2012}. Without a gas cell in front of the spectrograph, the calculation can be performed directly by comparing the observed spectra. TERRA first shifts and combines all high S/N observed spectra.  Their mean shift is the reference velocity for the rest of the observations.  Precise velocities come from a Gaussian fit to the minimum of the cross-correlation function between the template and individual observations. We correct this for internal wavelength drift, measured by a ThAr gas lamp spectrum taken at the same time as the observations. Internal velocity precision for Proxima is $\sim$0.9 m/s.

\subsection{Radial Velocity Constraints}\label{rvs}

In order to analyse the latest RV data for this work, which had the aim of confirming the existence of Proxima b and searching for additional companions on orbits interior to that of the planet, we employed the Exoplanet Mcmc Parallel tEmpering Radial VelOcity fitteR (EMPEROR) code \citep{PenaEtal2019}. The algorithm uses MCMC to explore the posterior parameter space, along with Bayesian statistics to determine if any signal exists.  In this work we employed EMPEROR with a first-order moving-average (MA), correlated-noise model, to smooth out the high-frequency noise that tends to dominate RV measurements.  No linear correlation terms were included, therefore we did not model any impact from stellar activity that is tracked by measured indices drawn from the stellar spectra themselves \citep[for example, see][]{diaz18}, beyond the MA model, since when included they were mostly found to be statistically similar to zero.  The model ($m(t)$) we employ as function of time for a given planet ($k$) and dataset ($d$) is described by  

\begin{equation}
    m(t) = \sum_{i=1}^{k} \sum_{j=1}^{d} [K_{i,j}(cos(\omega_{i,j} + T_{i,j}(t)) + e_{i,j}cos\omega_{i,j})]     
    + \dot{\gamma} + \sigma_{jit,j} + MA_{j}~~~~~
    \label{eq:rvmod}
\end{equation}

where $K$ is the semiamplitude of the planet model, $e$ is the eccentricity, $\omega$ is longitude of periastron, $T$ corresponds to the time of periastron passage, $\dot{\gamma}$ is the linear trend added to the model, $\sigma_{jit}$ is the excess jitter noise, and $MA$ is the moving average model.

The full timeseries was modeled as four separate datasets simultaneously, those data coming from four separate programs, three using HARPS and one using UVES (see Table~\ref{tab:data} for a brief summary).  The pre-2016 HARPS data, HARPS Pale Red Dot (PRD), and UVES data were discussed in \citep{AngladaEscudeEtal2016natProxCenb} and the HARPS Red Dots (RD) data, which was an extension and expansion of the PRD program and followed the observing procedure set out there, can be found at the website\footnote{https://reddots.space/}.  The simultaneous modeling contained five Keplerian parameters to model the planet, along with independent data offsets, MA coefficients, and excess noise (jitter) parameters, and finally a linear trend (Equation~\ref{eq:rvmod}). The MA modeling finds all four datasets have correlation coefficients that are statistically significantly different from zero at around the 3$\sigma$ level of confidence (see Table~\ref{tab:rvs}).  Therefore, both the high cadence and low cadence data require a correlated noise model to extract the most RV information from the negative effects of the noise.  It also appears that the jitter level slightly decreases between the PRD timeseries and the RD data, possibly due to a decrease in the activity state of Proxima.  The pre-2016 and UVES jitter values are significantly larger, likely due to the lower precision of these datasets compared to the post-HARPS upgrade observations.  No signficant linear trend was found in the full timeseries, however this does not rule out longer period companions as we split the data up into individual runs and we also nightly binned any data that were observed on the same night.  Such data handling would ultimately disfavour long period signals in the data, which is fine for these purposes since we were looking to constrain any planets with orbital periods less than that of Proxima b.  

\begin{table*}[t]
	\centering
	\caption{Summary of the RV data sets.}\label{tab:data}
	\begin{tabular}{lcccc} 
    \hline
Data set  &  Instrument & No. Observations & Baseline & Cadence \\
\hline
UVES   &  UVES   &  77 & 31/03/2000 $\--$ 30/03/2007 & Low \\
pre-2016 &  HARPS  &  63 &  27/05/2004 $\--$ 23/03/2013 & Low \\
PRD    &  HARPS  &  53 & 19/01/2016 $\--$ 01/04/2016 & High \\
RD     &  HARPS  &  62 & 01/06/2017 $\--$ 30/09/2017 & High \\
\hline
\end{tabular}
\end{table*}

\begin{figure}
	\hspace{-0.5cm}\vspace{-0.5cm}\includegraphics[width=10cm]{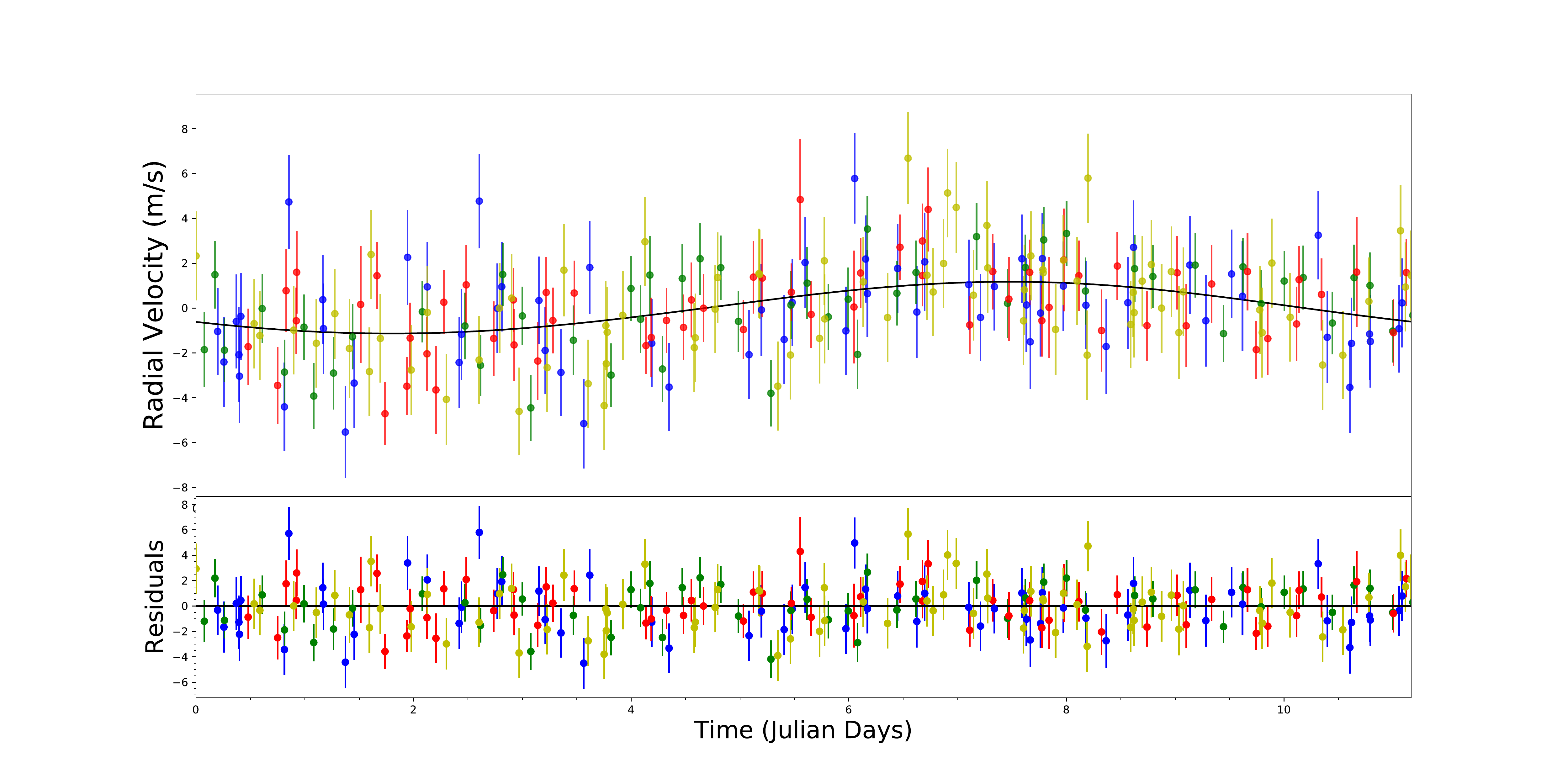}
    \caption{{\bf Top:} RV measurements of Proxima Centauri from HARPS prior to the Pale Red Dot program (blue), HARPS Pale Red Dot (green), HARPS Red Dots (red), and UVES (yellow), phase-folded to the planet's orbital period.  The black line is the best-fit Keplerian model.  {\bf Bottom:} Residuals to the fit.}
    \label{fig:rvs}
\end{figure}

EMPEROR provides excellent constraints on the orbital characteristics of Proxima b, particularly refining the orbital period of 11.1855~days to a level better than $\pm$2.3~minutes (see Table~\ref{tab:rvs}).  The eccentricity is also better constrained, with a 3$\sigma$ upper limit of 0.29, a movement towards zero of 0.06 compared with the value published in \citep{AngladaEscudeEtal2016natProxCenb}.  Further limits on the eccentricity are warranted, since lower values of eccentricity require the planet's orbit to be tidally locked to the star, providing additional constraints on the habitability of Proxima b \citep[see for example,][]{ribas16}.  These results highlight that the latest HARPS data are in excellent agreement with the previous data (see Fig.~\ref{fig:rvs}) and strongly confirm the existence of Proxima b. We find that at orbital periods shorter than that of Proxima b, the radial-velocity precision we can reach is $\sim$0.5~m/s, mainly coming from the high-cadence datasets, placing an upper limit of $\sim$0.5~\me\, on any possible inner Proxima c.  The semiamplitude of the signal we find here is also in excellent agreement with that already found for Proxima~b, with a difference of only 0.06~\ms\, between the model published in \citet{AngladaEscudeEtal2016natProxCenb} and the model found here; although the uncertainties we find are almost half those found previously.  

\begin{table}
	\centering
	\caption{Orbital constraints and nuisance parameters for the Proxima b model from the EMPEROR analysis of RV data.}\label{tab:rvs}
	\begin{tabular}{lc} 
    \hline
Orbital Parameter  &  Model Value \\
\hline
Amplitude [\ms] &   $1.32_{-0.14}^{+0.12}$ \\
Period [d] &   $11.1855_{-0.0014}^{+0.0016}$ \\
Phase [rads]    & $3.44_{-1.79}^{+0.62}$ \\
Longitude [rads]     &   $4.40_{-3.47}^{+0.84}$ \\
Eccentricity    &   $0.08_{-0.06}^{+0.07}$ \\
$\dot{\gamma}$ [ms$^{-1}$d$^{-1}$] & $-0.0005_{-0.0002}^{+0.0003}$ \\
$\sigma_{jit,pre-2016}$ [\ms]  &   $1.84_{-0.11}^{+0.09}$ \\
$\gamma_{pre-2016}$ [\ms]  &  $1.23_{-0.73}^{+0.67}$ \\
$\phi_{pre-2016}$       & $0.63_{-0.11}^{+0.16}$ \\
$\tau_{pre-2016}$ [d] & $7.57_{-3.10}^{+1.52}$ \\
$\sigma_{jit,PRD}$ [\ms]  &   $1.44_{-0.20}^{+0.10}$ \\
$\gamma_{PRD}$ [\ms]  &  $1.98_{-0.92}^{+1.08}$ \\
$\phi_{PRD}$       & $0.38_{-0.11}^{+0.17}$ \\
$\tau_{PRD}$ [d] & $7.86_{-4.46}^{+0.97}$ \\
$\sigma_{jit,RD}$ [\ms]  &  $1.14_{-0.10}^{+0.13}$ \\
$\gamma_{RD}$ [\ms]  &   $2.10_{-0.90}^{+1.21}$ \\
$\phi_{RD}$     &  $0.50_{-0.20}^{+0.16}$ \\
$\tau_{RD}$ [d] &  $6.91_{-2.58}^{+2.33}$ \\
$\sigma_{jit,UVES}$ [\ms]  &  $1.91_{-0.11}^{+0.09}$ \\
$\gamma_{UVES}$ [\ms] &  $-0.20_{-0.32}^{+0.27}$ \\
$\phi_{UVES}$     & $0.62_{-0.17}^{+0.19}$ \\
$\tau_{UVES}$ [d] & $4.54_{-1.63}^{+2.35}$ \\
\hline
	\end{tabular}
	\linebreak 
	pre-2016 - parameters for data taken with HARPS prior to the PRD program \linebreak
	PRD - parameters for the Pale Red Dot program  \linebreak
	RD - parameters for the Red Dots program
\end{table}

\section{Summary}\label{conclusions}

We have addressed the recent claims of transit-like events in optical photometry arising from the habitable-zone terrestrial planet Proxima~b.  We observed the system with the Spitzer Space Telescope for $\sim$48~hours at 4.5~{\microns}. The observations covered the 99\% probability window predicted for the transit using the published RV model in \citet{AngladaEscudeEtal2016natProxCenb}.  The limits on this window were drawn from the posterior density distribution of the model, assuming 99\% uncertainty limits on the model parameters like period and eccentricity.

Our observations and BLISS analysis allowed us to reach an unbinned photometric precision of 7500~ppm, with a 2~hr (rough transit duration) binned precision of 200~ppm.  No transit-like event could be attributed to the passage of Proxima~b in front of its star. The previously witnessed transit-like events may result from residual correlated noise arising from the star's complex and frequent flaring and activity patterns.  Our photometric precision places a 3$\sigma$ upper limit on the size of a transiting Proxima~b of 0.4~\re.  This corresponds to an implausible minimum density of $\sim$112 g cm$^{-3}$.  

We performed a short radial-velocity experiment to search for additional small planets interior to the orbit of Proxima b, whilst constraining better Proxima b's orbital characteristics.  Beyond the data published in \citet{AngladaEscudeEtal2016natProxCenb}, we also included newly observed HARPS data from the Red Dots program.  After fitting for the orbit of Proxima~b, the residuals reveal no inner planet down to the 0.5~m/s level, which relates to planets with minimum masses of 0.5~\me.  The orbital period and eccentricity of Proxima b's orbital solution were also better constrained in this process, with a precision in period of better than $\pm$30~s found, and a 3$\sigma$ upper limit on the eccentricity of 0.29.

Finally, we did witness a transit-like event at the 0.3\% depth level and with an asymmetric morphology.  However, we found we could remove the feature completely from the time series by using variable-radius photometry apertures. A study of this and similar features in additional Spitzer data of Proxima and beyond, as well as detection and treatment methods, will be published in a future paper (Challener et al. 2019, in prep).

\section*{Acknowledgments}

We thank the Spitzer Science Center staff for making these observations 
possible. This work
is based on observations made with the {\em Spitzer Space Telescope},
which is operated by the Jet Propulsion Laboratory, California
Institute of Technology under a contract with NASA.  We also thank the anonymous referee for their efficient and detailed review.  The authors acknowledge support from the following:
CATA-Basal/Chile PB06 Conicyt and Fondecyt/Chile project \#1161218 (JSJ).
CONICYT Chile through CONICYT-PFCHA/Doctorado Nacional/2017-21171752 (JP).
Spanish MINECO programs AYA2016-79245-C03-03-P, ESP2017-87676-C05-02-R (ER), ESP2016-80435-C2-2-R (EP) and through the "Centre of Excellence Severo Ochoa" award SEV-2017-0709 (PJA, CRL and ER).  
STFC Consolidated Grant ST/P000592/1 (GAE). 
NASA Planetary Atmospheres Program grant NNX12AI69G, NASA Astrophysics Data Analysis Program grant NNX13AF38G.  
Spanish Ministry of Science, Innovation and Universities and the Fondo Europeo de Desarrollo Regional (FEDER) through grant ESP2016-80435-C2-1-R (IR). We thank contributors to SciPy, Matplotlib, and the Python Programming
Language; the free and open-source community; and the NASA
Astrophysics Data System for software and services.
Based on observations made with ESO Telescopes at the La Silla Paranal Observatory under programmes 096.C-0082, 191.C-0505, and 099.C-0880.

\bibliographystyle{mnras}
\bibliography{refs}

\label{lastpage}

\end{document}